\documentclass[11pt, prd, aps, showpacs, nofootinbib, superscriptaddress]{revtex4-1}
\usepackage{graphicx}
\usepackage{dcolumn}
\usepackage{bm}
\usepackage{amsmath}
\usepackage{comment}
\usepackage[utf8]{inputenc}
\usepackage{physics}
\usepackage{xcolor}
\newcommand{\bea}{\begin{eqnarray}}
\newcommand{\eea}{\end{eqnarray}}
\usepackage{mathtools}
\usepackage{cancel} 
\newcommand{\monec}{\raisebox{-0.43\totalheight}{\includegraphics[scale=0.55]{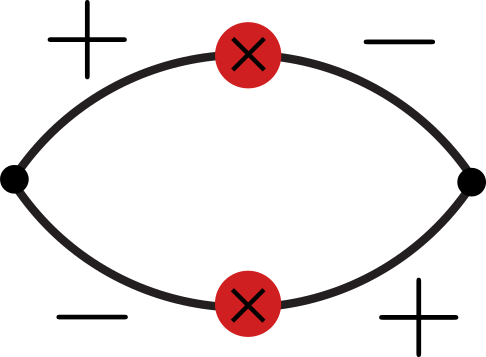}}}
\newcommand{\moned}{\raisebox{-0.43\totalheight}{\includegraphics[scale=0.55]{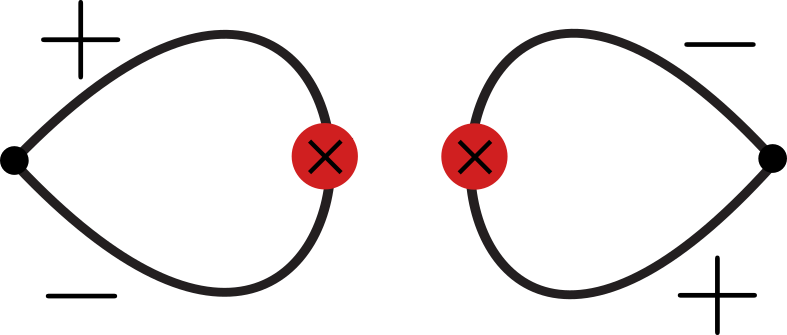}}}
\newcommand{\mtwoc}{\raisebox{-0.43\totalheight}{\includegraphics[scale=0.55]{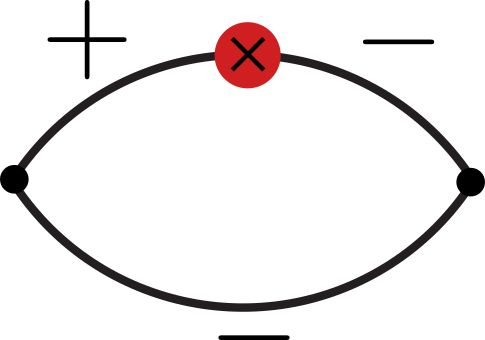}}}
\newcommand{\mtwod}{\raisebox{-0.43\totalheight}{\includegraphics[scale=0.55]{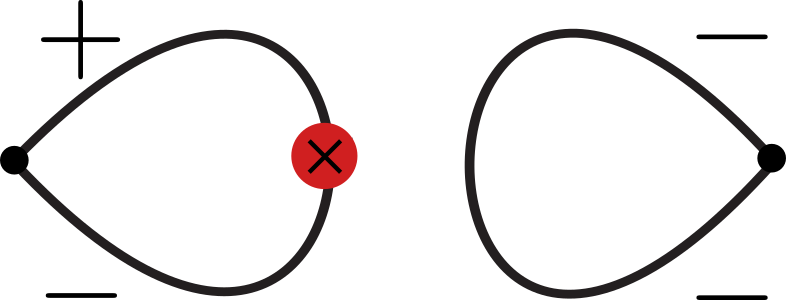}}}

\usepackage{hyperref}
\hypersetup{
    colorlinks=true,
    linkcolor=blue,
    filecolor=magenta,      
    urlcolor=cyan,
}
\usepackage[all]{hypcap}

\begin{document}

\title{First direct lattice calculation of the chiral perturbation theory low-energy constant $\ell_7$.}
\author{R. Frezzotti}
\email{roberto.frezzotti@roma2.infn.it}
\affiliation{Dipartimento di Fisica and INFN, Università di Roma “Tor Vergata”,
Via della Ricerca Scientifica 1, I-00133 Rome, Italy}
\author{G. Gagliardi} 
\email{giuseppe.gagliardi@roma3.infn.it}
\affiliation{Istituto Nazionale di Fisica Nucleare, Sezione di Roma Tre,
Via della Vasca Navale 84, I-00146 Rome, Italy}
\author{V. Lubicz}
\email{vittorio.lubicz@uniroma3.it}
\affiliation{Dipartimento di Fisica, Università Roma Tre and INFN, Sezione di Roma Tre,
Via della Vasca Navale 84, I-00146 Rome, Italy}
\author{G. Martinelli}
\email{guido.martinelli@roma1.infn.it}
\affiliation{Dipartimento di Fisica and INFN Sezione di Roma La Sapienza, Piazzale Aldo Moro 5, I-00185 Rome, Italy}
\author{F. Sanfilippo}
\email{francesco.sanfilippo@infn.it}
\affiliation{Istituto Nazionale di Fisica Nucleare, Sezione di Roma Tre,
Via della Vasca Navale 84, I-00146 Rome, Italy}
\author{S. Simula}
\email{silvano.simula@roma3.infn.it}
\affiliation{Istituto Nazionale di Fisica Nucleare, Sezione di Roma Tre,
Via della Vasca Navale 84, I-00146 Rome, Italy}
\date{\today}

\begin{abstract}
We evaluate by means of lattice QCD calculations the low-energy constant $\ell_{7}$ which parametrizes strong isospin effects at NLO in $\rm{SU}(2)$ chiral perturbation theory. Among all low-energy constants at NLO, $\ell_{7}$ is the one known less precisely, and its uncertainty is currently larger than $50\%$. Our strategy is based on the RM123 approach in which the lattice path-integral is expanded in powers of the isospin breaking parameter $\Delta m= (m_{d}-m_{u})/2$. In order to evaluate the relevant lattice correlators we make use of the recently proposed \textit{rotated twisted-mass} (RTM) scheme. Within the RM123 approach, it is possible to cleanly extract the value of $\ell_{7}$ from either the pion mass splitting $M_{\pi^{+}}-M_{\pi^{0}}$ induced by strong isospin breaking at order $\mathcal{O}\left((\Delta m)^{2}\right)$ (mass method), or from the coupling of the neutral pion $\pi^{0}$ to the isoscalar operator  $\left(\bar{u}\gamma_{5}u + \bar{d}\gamma_{5} d\right)/\sqrt{2}$ at order $\mathcal{O}(\Delta m)$ (matrix element method). In this pilot study we limit the analysis to a single ensemble generated by the Extended Twisted Mass Collaboration (ETMC) with $N_{f}=2+1+1$ dynamical quark flavours, which corresponds to a lattice spacing $a\simeq 0.095~{\rm fm}$ and to a pion mass $M_{\pi}\simeq  260~{\rm MeV}$. We find that the matrix element method outperforms the mass method in terms of resulting statistical accuracy. Our determination, $\ell_{7} = 2.5(1.4)\times 10^{-3}$, is in agreement and improves previous calculations.  
\end{abstract}

\maketitle

\section{Introduction}
Chiral Perturbation Theory (ChPT) represents a powerful theoretical framework to describe the low-energy dynamics of QCD taking full advantage of the consequences of spontaneous chiral symmetry breaking. The ChPT Lagrangian is organized as a power expansion in terms of the external momenta and quark masses and, depending on whether only light quarks are considered or the strange quark is included, one has $\textrm{SU}(2)$ or $\text{SU}(3)$ ChPT. The chiral expansion is then written in terms of low-energy constants (LECs) whose values are fixed by matching a number of observables to the predictions of fundamental QCD or to their experimental determination. The ChPT Lagrangian at LO contains only two LECs: the pion decay constant $f_{\pi}\sim 132~{\rm MeV}$ and the parameter $B_{0}$ proportional to the chiral condensate $\langle \bar{\psi}\psi\rangle$, while at NLO the $\rm{SU}(2)$ ChPT Lagrangian is parametrized, apart from contact terms,  by the seven LECs $\{\ell_{i}\}_{i=1,\ldots,7}$~\cite{Gasser:1983yg}. Strong isospin-breaking (IB) effects at NLO are all parametrized by $\ell_{7}$, which corresponds to the only LEC that couples to an operator function of the up-down quark mass difference $m_{u}-m_{d}$. \
In the seminal papers by Gasser and Leutwyler~\cite{Gasser:1983yg, GASSER1985465}, a phenomenological estimate of the values of the NLO LECs $\{\ell_{i}\}_{i=1,\ldots,7}$, based on the available experimental informations, was given.\\

$\ell_{7}$ enters any ChPT-based calculations where IB effects play an important role, and for this reason a first principle and high precision  evaluation of its value is important for several phenomenological analyses. To give some example, according to Ref.~\cite{di_Cortona_2016} the prediction at NLO of the axion mass $m_{a}=5.70(6)(4)~{\rm \mu eV}\left( 10^{12}~{\rm GeV}/f_{a}\right)$, where $f_{a}$ is the axion decay constant, has an overall uncertainty of order $\mathcal{O}(1\%)$ where the first source of error comes from the uncertainty on the up-down quark mass ratio $m_{u}/m_{d}$, while the second one is due to the uncertainty on the value of $\ell_{7}$. A precise determination of $\ell_{7}$ turns out to be even more crucial in the evaluation at NLO of the axion quartic self-coupling $\lambda_{a} = -0.346(22)~m_{a}^{2}/f_{a}^{2}$, where the resulting $\simeq 6\%$ error is completely dominated by the uncertainty on $\ell_{7}$. More recently~\cite{diluzio2021axion}, the axion-pion scattering process $a \pi \to \pi \pi$ has been computed in $\rm{SU}(2)$ ChPT at NLO in order to probe the convergence of the chiral expansion of the axion-pion thermalization rate $\Gamma_{a}(T)$, from which it is possible to put the so-called hot dark matter (HDM) bounds on the axion mass. In this case, the uncertainty on the value of $\ell_{7}$ produces a $15-20\%$ uncertainty in the amplitude for $a \pi \to \pi \pi$ at NLO. \\

Among the LECs $\ell_{7}$ turns out to be one affected by the largest uncertainty. A first estimate of its value was obtained by matching the charged/neutral pion mass splitting $M_{\pi^{+}}-M_{\pi^{0}}$ parametrized by $\ell_{7}$ in $\rm{SU}(2)$ ChPT at NLO to the value predicted by $\rm{SU}(3)$ ChPT at LO and due to $\eta-\pi^{0}$ mixing~\cite{Gasser:1983yg}. This gives, for this constant which is scheme and scale independent at NLO, the relation
\begin{align}
\label{eq:LO_SU3_pred}
\ell_{7} = \frac{f_{\pi}^{2}}{12 M_{\eta}^{2}} \sim 5 \times 10^{-3}~.
\end{align}
Higher order corrections to the $\rm{SU}(3)$ tree level prediction, which depend on the knowledge of the $\rm{SU}(3)$ LECs at NLO, however, turn out to be numerically of the same size as the LO prediction of Eq.~(\ref{eq:LO_SU3_pred}), giving rise to a large systematic uncertainty, namely $\ell_{7} = 7(4)\times 10^{-3}$~\cite{di_Cortona_2016}. 
A first attempt to determine the value of $\ell_{7}$ from lattice QCD simulations has been made by the RBC-UKQCD Collaboration~\cite{Boyle_2016}, where the NLO and NNLO partially quenched $\rm{SU}(2)$ LECs have been extracted by means of a global fit to various pseudoscalar masses and decay constants. However, the value $\ell_{7}= 6.5(3.8) \times 10^{-3}$ that has been reported, being obtained from a NNLO ChPT global fit, is strongly correlated to the other LECs which contribute at NLO to the dependence of the fitted meson masses and decay constants on the light quark masses. \\

In this paper we propose to determine $\ell_{7}$ directly from first principle lattice QCD simulations, evaluating IB effects within the RM123 approach~\cite{de_Divitiis_2012, Giusti:2017dmp}, in which the path-integral is expanded around the isosymmetric point $m_{d}=m_{u}$ in powers of $\Delta m= \left(m_{d}-m_{u}\right)/2$. We use the \textit{rotated twisted-mass} (RTM) scheme~\cite{Fr:2021}, which have been shown to reduce the statistical noise of some mesonic lattice correlation functions.
We employ two different strategies to determine $\ell_{7}$: the first one (in the following \textit{mass method}) is based on the computation of the charged/neutral pion mass splitting $M_{\pi^{+}}-M_{\pi^{0}}$ at order $\mathcal{O}\left((\Delta m)^2 \right)$, while the second  one (in the following \textit{matrix element method}) consists in extracting $\ell_{7}$ from the coupling $Z_{P^{0}\pi^{0}}=\langle 0 | P^{0} | \pi^{0}\rangle$ of the neutral pion to the isoscalar operator 
\begin{align}
P^{0} = \frac{1}{\sqrt{2}}\left(\bar{u}\gamma_{5}u + \bar{d}\gamma_{5}d\right)
\end{align}
at leading order $\mathcal{O}(\Delta m)$. The clear advantage of our strategy is represented by the fact that within the RM123 approach one evaluates directly the derivatives in $\Delta m$ of both $M_{\pi^{+}}-M_{\pi^{0}}$ and $Z_{P^{0}\pi^{0}}$ which, being proportional to $\ell_{7}$, allow for a clean extraction of its value. This is different from what happens in the global fit procedure, where the value of $\ell_{7}$ must be evaluated together with the other LECs which enter at the same and lower order in ChPT, by fitting the combined quark mass dependence of several meson masses and decay constants.\\

We make use of the gauge configurations produced with Wilson-clover TM fermions by the Extended Twisted Mass Collaboration (ETMC)~\cite{alexandrou2021ratio,alexandrou2021quark}. For this feasibility study, we limit our simulations to a single value of the lattice spacing $a\simeq 0.095~{\rm fm}$ and to an higher-than-physical pion mass $M_{\pi}\simeq  260~{\rm MeV}$, postponing the extrapolation to the continuum and chiral limit to a future work. We also note, however, that any residual pion mass dependence which is left in our present estimate of $\ell_{7}$ represents a NNLO (or higher order) effect in ChPT. This is only relevant for phenomenological applications which aim to an accuracy beyond NLO in the chiral expansion, where many other unknown LECs are involved in any case.  \\

The remaining of the paper is organized as follows: in Sec.~[\ref{sec:strategy}] we briefly introduce the mass method and the matrix element method. In Sec.~[\ref{sec:RM_123_RTM}] we derive the diagrammatic expansion of the relevant lattice correlators in the RTM scheme, describing the procedure we used to relate them to the pion mass splitting $M_{\pi^{+}}-M_{\pi^{0}}$ and to the matrix element $Z_{P^{0}\pi^{0}}$. In Sec.~[\ref{sec:Num_result}] we present our numerical results, and finally in Sec.~[\ref{sec:Conclusions}] we draw our conclusions.

\section{The mass method and the matrix element method}
\label{sec:strategy}
In this section we describe the two methods that, following Gasser and Leutwyler~\cite{Gasser:1983yg}, we considered for a direct determination of $\ell_{7}$. 
The mass method relies on the fact that $\ell_{7}$ parametrizes the charged/neutral pion mass difference induced by QCD IB through 
\begin{equation}
\label{eq:mpi_l7}
M_{\pi^+}^2-M_{\pi^0}^2=\left(m_u-m_d\right)^2\frac{4B_{0}^2}{f_{\pi}^2}\,\ell_7\,,\\[7pt]
\end{equation}
where $f_{\pi}$ is the pion decay constant normalized as $f_{\pi} \simeq 132~{\rm MeV}$. Expanding the l.h.s. of the previous equation using 
\begin{equation}
    M_{\pi^+}^2-M_{\pi^0}^2= \left(M_{\pi^{+}}+M_{\pi^{0}}\right)\cdot\left( M_{\pi^{+}}- M_{\pi^{0}}\right)\simeq 2M_\pi\left(M_{\pi_+}-M_{\pi_0}\right)\,,\\[7pt]
\end{equation}
and by noticing that at LO in the chiral expansion $M_\pi^2\simeq B_{0}\left(m_u+m_d\right)=2B_{0}m_l$, one has 
\begin{equation}
    2M_\pi \left(M_{\pi^+}-M_{\pi^0}\right)\simeq (m_{u}-m_{d})^{2}\frac{M_\pi^4}{m_l^2 f_{\pi}^2}\,\ell_7\,.\\[7pt]
\end{equation}
This allows to compute $\ell_{7}$ through
\begin{equation}
\label{l7_M1_definition}
    \ell_7=2\frac{\left(M_{\pi^+}-M_{\pi^0}\right)_{QCD}}{(m_{u}-m_{d})^{2}}\cdot\frac{m_l^2 f_{\pi}^2}{M_\pi^3}\,,
\end{equation}
where we emphasized that the difference $\left( M_{\pi^{+}}- M_{\pi^{0}}\right)_{QCD}$ in Eq.~(\ref{l7_M1_definition}) indicates only the pure QCD contribution to the pion mass splitting, which is subdominant with respect to the leading QED contribution of $\mathcal{O}(\alpha_{em})$.\\

The matrix element method relies instead on the fact that, away from the isosymmetric limit, i.e. for different up and down quark masses, the neutral pion has a non vanishing iso-singlet component the size of which is quantified by the matrix element
\begin{align}
Z_{P^{0}\pi^{0}} &\equiv \langle 0 | P^{0} | \pi^{0}\rangle =  \frac{1}{\sqrt{2}}\langle 0 | \left( \bar{u}\gamma^{5}u + \bar{d}\gamma^{5}d\right) | \pi^{0} \rangle
\end{align}
The matrix element $Z_{P^{0}\pi^{0}}$ is directly proportional to $\ell_{7}$ through~\cite{Gasser:1983yg}
\begin{align}
\label{eq:Zp0pi_l7}
Z_{P^{0}\pi^{0}}= -(m_{u}-m_{d}) \frac{4B_{0}^{2}}{f_{\pi}}\,\ell_{7} = -(m_{u}-m_{d})\frac{M_{\pi}^{4}}{f_{\pi}m_{\ell}^{2}}\,\ell_{7}~,
\end{align}
which allows to determine $\ell_{7}$ via
\begin{align}
\label{l7_M2_definition}
\ell_{7} = -\frac{Z_{P^{0}\pi^{0}}}{m_{u}-m_{d}}\cdot \frac{f_{\pi}m_{\ell}^{2}}{M_{\pi}^{4}}~.    
\end{align}
Eqs.~(\ref{l7_M1_definition}) and~(\ref{l7_M2_definition}) show that in a mass independent scheme $\ell_{7}$ is a dimensionless RGI quantity. Therefore it can be expressed equivalently in terms of the bare lattice quantities or the renormalized  ones.\\

To evaluate the pion mass splitting at order $\mathcal{O}\left( (m_{u}-m_{d})^{2}\right)$ and the matrix element $Z_{P^{0}\pi^{0}}$ at $\mathcal{O}\left(m_{u}-m_{d}\right)$, we adopt the RM123 method which is based on the Taylor expansion of the QCD path-integral around the isosymmetric point~\cite{de_Divitiis_2012,Giusti:2017dmp}. In order to reduce the statistical noise of the correlators involved in the calculation, we will make use of the RTM scheme introduced in Ref.~\cite{Fr:2021}. For completeness this scheme will be briefly introduced in the next section, the interested reader is referred to Ref.~\cite{Fr:2021} for more details.


\section{RM123 expansion in the RTM scheme}
\label{sec:RM_123_RTM}
The lattice QCD RTM Lagrangian of the light doublet $\psi'_{\ell} = (u',d')$, is given by~\cite{Fr:2021}
\begin{align}
\label{RTM}
\mathcal{L}_{RTM}(\psi'_{\ell}) = \bar{\psi}'_{\ell}(x)\left[ \gamma_{\mu}\widetilde{\nabla}_{\mu}  -i\gamma_{5}\tau_{3}W(m_{cr})+ m_{\ell} + \Delta m\tau_{1}   \right]\psi'_{\ell}(x)~,
\end{align}
where $\Delta m = \frac{1}{2}(m_{d}-m_{u})$, $\widetilde \nabla_\mu$ is the lattice symmetric covariant derivative, written in terms of the forward $(\nabla_\mu )$ and backward ($ \nabla^*_\mu$) covariant derivatives, 
\begin{equation}
\widetilde{\nabla}_\mu = \frac{1}{2} \left( \nabla^*_\mu + \nabla_\mu \right)
\end{equation}
and $W(m_{cr})$ is the critical Wilson term, which includes the mass and is globally odd under $r \to -r$,
\begin{equation}
W(m_{cr}) = - a\, \frac{r}{2}\, \nabla_\mu \nabla^*_\mu + m_{cr}(r)\, .
\end{equation}
Notice the unconventional direction in flavour space of the isospin-breaking term
\begin{align}
\mathcal{L}_{IB}=\bar{\psi}'_{\ell} \tau_{1}\psi'_{\ell}~.
\end{align}
The quark fields $u',d'$ appearing in the RTM Lagrangian, which are regularized in Eq.~(\ref{RTM}) with opposite values of the Wilson parameter $r=\pm 1$,  are not the physical ones. They are related to the physical up and down quark fields $u$ and $d$, through
\begin{align}
\label{eq:primed_to_phys}
\begin{pmatrix}
u' \\
d' 
\end{pmatrix}
= \frac{1}{\sqrt{2}}\begin{pmatrix}
1 & 1 \\
-1 & 1
\end{pmatrix}
\begin{pmatrix}
u \\
d 
\end{pmatrix}
= \frac{1}{\sqrt{2}}\begin{pmatrix}
u+d \\
d-u
\end{pmatrix}
\end{align}\\[6pt]
The RTM Lagrangian is not equivalent to the standard twisted mass Lagrangian. It can be shown~\cite{Fr:2021} that rotating back to the physical doublet $\psi_{\ell}=(u,d)$, the RTM Lagrangian coincides with the regularization proposed in Ref.~\cite{Frezzotti_2004} and adopted by the ETMC to discretize the Lagrangian of the heavy doublet $\psi_{h}=(c,s)$. From this observation, it follows that the Lagrangian of Eq.~(\ref{RTM}) inherits all the benefits of the standard twisted mass regularization, including the $\mathcal{O}(a^{2})$ improvement of parity even observables at maximal twist.\\

In order to determine the diagrammatic expansion (in the RTM basis)  of $M_{\pi^{+}}-M_{\pi^{0}}$ at order $\mathcal{O}(\Delta m^{2})$ and the matrix element $Z_{P^{0}\pi^{0}}$ at order $\mathcal{O}(\Delta m)$ it is necessary to establish the relation between the physical correlators $C_{\pi^{+}\pi^{+}}(t) - C_{\pi^{0}\pi^{0}}(t)$ and $C_{P^{0}\pi^{0}}(t)$ ($C_{AB}(t) = \langle 0 | A(t) B^{\dag}(0) | 0 \rangle)$, written in terms of the physical quark fields $u$ and $d$,  and the correlators written in the rotated basis. Using Eq.~(\ref{eq:primed_to_phys}), it is straightforward to show that such relations are given by
\begin{align}
\label{C1}
&C_{\pi^{+}\pi^{+}}(t) - C_{\pi^{0}\pi^{0}}(t) = -2\,C_{\pi'^{+}\pi'^{-}}(t)~, \\[8pt]
\label{C2}
&C_{P^{0}\pi^{0}}(t) =  -\frac{1}{\sqrt{2}}\left[ C_{P'^{0}\pi'^{+}}(t) + C_{P'^{0}\pi'^{-}}(t)\right] ~,
\end{align}
where 
\begin{align}
\pi'^{-} = \bar{u}'\gamma_{5}d', \qquad \pi'^{+}= \bar{d}'\gamma_{5}u', \qquad P'^{0} = \frac{1}{\sqrt{2}}\left[\bar{u}'\gamma_{5} u + \bar{d}'\gamma_{5} d\right]~.    
\end{align}
\newpage
We now discuss the RM123 expansion of the correlators appearing in the r.h.s. of Eqs.~(\ref{C1}) and~(\ref{C2}), respectively at second and first order in $\Delta m$, postponing to the next subsection the description of the relations between the correlators and the physical observables that we want to extract. \\

In full generality, the expansion of the VEV of a given observable $\mathcal{O}$ up to second order in $\Delta m$, can be written as
\begin{align}
\langle \mathcal{O}\rangle &= \langle \mathcal{O}\rangle_{0}  -\Delta m\sum_{x}\langle \mathcal{O}\mathcal{L}_{IB}(x)\rangle_{0} +\nonumber \\[10pt]
&+\frac{(\Delta m)^{2}}{2}\sum_{x,y}\left[\langle\mathcal{O}\mathcal{L}_{IB}(x)\mathcal{L}_{IB}(y)\rangle_{0} - \langle \mathcal{O}\rangle_{0}\langle \mathcal{L}_{IB}(x)\mathcal{L}_{IB}(y)\rangle_{0}\right] + \ldots~,
\end{align}
where $\langle \cdot \rangle_{0}$ denotes the average in isosymmetric QCD. For the correlators $C_{\pi'^{+}\pi'^{-}}(t)$, $C_{P'^{0}\pi'^{+}}(t)$ and $C_{P'^{0}\pi'^{-}}(t)$, the expansion reads (the latter two must be expanded at leading order only)
\begin{align}
C_{\pi'^{+}\pi'^{-}}(t) &=  \frac{(\Delta m)^{2}}{2}\sum_{x,y}\left[\langle  P_{\pi'^{+}}(t)\mathcal{L}_{IB}(x)\mathcal{L}_{IB}(y)P^{\dag}_{\pi'^{-}}(0)\rangle_{0}\right]  \\[10pt]
C_{P'^{0}\pi'^{+}}(t) &= -\Delta m  \sum_{x}\langle P'^{0 }(t)\mathcal{L}_{IB}(x)P^{\dag}_{\pi'^{+}}(0)\rangle_{0}  \\[10pt]
C_{P'^{0}\pi'^{-}}(t) &=  -\Delta m  \sum_{x}\langle P'^{0}(t)\mathcal{L}_{IB}(x)P^{\dag}_{\pi'^{-}}(0)\rangle_{0}~,
\end{align}
where $P_{P}(x)$ is an interpolator of the meson $P$. In the previous expression it is implied that the interpolating fields are projected to zero three momentum $\vec{p}=0$. In all cases, the leading zeroth order term vanishes because in the isosymmetric limit there is no mixing between the rotated pions, and they also do not possess an isoscalar component. Moreover, the order $\mathcal{O}(\Delta m)$ term in the expansion of $C_{\pi'^{+}\pi'^{-}}$ also vanishes (as expected), because at least two insertions of the perturbation $\mathcal{L}_{IB}$ are needed in order to convert $\bar{u}'\leftrightarrow \bar{d}'$ and have a mixing between $\pi'^{+}$ and $\pi'^{-}$. In the physical basis, this corresponds to the fact that the pion correlators are symmetric with respect to $u\leftrightarrow d$ and can receive, therefore, only corrections proportional to even powers of $\Delta m$. \\

Performing the corresponding Wick contractions and then taking the isosymmetric limit, one obtains the following diagrammatic expansion for the previous correlators\footnote{In Eqs.~(\ref{M1_expansion}) and~(\ref{M2_expansion}), the quark-line connected and disconnected Wick contractions have a relative minus sign stemming from the extra fermion loop present in the disconnected contribution. For later convenience, we decided to pull out this extra minus sign from the definition of the disconnected diagrams. } 
\begin{align}
\label{M1_expansion}
&\begin{aligned} 
C_{\pi^{+}\pi^{+}}(t) - C_{\pi^{0}\pi^{0}}(t)  &= -2\,C_{\pi'^{+}\pi'^{-}}(t)\\[12pt]
&= -2\left(\frac{Z_{S}}{Z_{P}}\right)^{2}(\Delta m)^{2}\left[\,\,\monec\,\, - \,\, \moned \,\,\right]\\[12pt]
&\equiv -2\left( \frac{Z_{S}}{Z_{P}}\right)^{2}(\Delta m)^{2}\left[ C_{MM}^{conn.}(t) - C_{MM}^{disc.}(t)\right]~,
\end{aligned} \\[18pt]
\label{M2_expansion}
&\begin{aligned} 
C_{P^{0}\pi^{0}}(t)  &= -\frac{1}{\sqrt{2}}\left( C_{P'^{0}\pi'^{+}}(t) + C_{P'^{0}\pi'^{-}}(t)\right) \\[12pt]
&= -2\frac{Z_{S}}{Z_{P}}\Delta m \left[\,\, \mtwoc\,\, - \,\, \mtwod\right] \\[12pt]
&\equiv -2\frac{Z_{S}}{Z_{P}}\Delta m \left[ C_{MEM}^{conn.}(t) - C_{MEM}^{disc.}(t)\right]~.
\end{aligned}
\end{align}
The black lines in the diagrams represent the isosymmetric light quark propagators with Wilson parameter $r=\pm 1$ as denoted in the plots, i.e. the isosymmetric propagator of the $u'$ and of the $d'$ quark. Black vertices denote the insertion of $\gamma_{5}$, while red vertices correspond to the insertion of the perturbation $\mathcal{L}_{IB}$. Diagrams obtained from one another through a simultaneous flip of the Wilson parameter of all  the propagators are equivalent. Finally, in Eqs.~(\ref{M1_expansion}),~(\ref{M2_expansion}) we included the renormalization constant (RC) of the operator $\mathcal{L}_{IB}$ and of the mass difference $m_{d}-m_{u} = 2\Delta m$ which, in our twisted mass formulation, are given respectively by the RC of the scalar density $Z_{S}$, and by the inverse of the RC of the pseudoscalar  density $Z_{P}^{-1}$.  \\

The main advantage of the RTM basis is that it allows to consider mesonic correlators where the quark and anti-quark fields entering the correlators are always discretized with opposite values of the Wilson parameter $r$, as shown in Eqs.~(\ref{M1_expansion}) and~(\ref{M2_expansion}). Such correlators are notoriously affected by a smaller statistical uncertainty w.r.t. correlators involving quark propagators with equal values of $r$.


\subsection{From correlators to the physical observables}
We now discuss how to relate the correlators defined in the previous section to the charged/neutral pion mass difference (mass method) and to the coupling $Z_{P^{0}\pi^{0}}$ of the neutral pion to the isoscalar density (matrix element method) induced by strong IB. \\

Let us start from the mass method, and consider the correlators of the charged and neutral pion in the physical basis. In the complete theory, where the perturbation $\Delta m\mathcal{L}_{IB}$ is treated to all orders, one has 
\begin{align}
C_{\pi^{+}\pi^{+}}(t) &= A_{\pi^{+}}(\Delta m)\cosh{\left[M_{\pi^{+}}(\Delta m) (T/2 -t )\right]} \,\, +\,\, \ldots\\[10pt]
C_{\pi^{0}\pi^{0}}(t) &= A_{\pi^{0}}(\Delta m)\cosh{\left[M_{\pi^{0}}(\Delta m) (T/2 -t )\right]}\,\, +\,\, \ldots
\end{align}
where $T$ is the temporal extent of the lattice, the dots indicate excited-state contributions which from now on will be neglected assuming the ground state dominance, and for a meson $P$ the amplitude $A_{P}$ is related to the matrix element $Z_{PP} = \langle P | P^{\dag}_{P}| 0\rangle$, through
\begin{align}
A_{P} = \frac{ |Z_{PP}|^{2}}{M_{P}}e^{-M_{P}T/2}~.
\end{align}
At fixed value of $m_{\ell}$, the masses $M_{\pi^{+}}, M_{\pi^{0}}$ and the amplitudes $A_{\pi^{+}}, A_{\pi^{0}}$ are implicitly functions of the quark mass difference. Hence we can express the derivatives of the correlators w.r.t. $\Delta m$ in terms of the derivatives of amplitudes and masses. For the generic correlator $C_{PP}(t)$ the first derivative in $\Delta m $ is given by $( Q^{'} = dQ/d(\Delta m))$
\begin{align}
C^{'}_{PP}(t) &= A'_{P}\cosh{\left[M_{P}(T/2-t)\right]} + A_{P}M'_{P}\cdot(T/2-t)\cdot\sinh{\left[ M_{P}(T/2-t)\right]}~,
\end{align}
while for the second derivative one gets
\begin{align}
C^{''}_{PP}(t) &= A''_{P}\cosh{\left[M_{P}(T/2-t)\right]} \nonumber \\[10pt]
&+ \left( 2A'_{P}M'_{P}
+ A_{P}M''_{P}\right)\cdot(T/2-t)\cdot\sinh{\left[ M_{P}(T/2-t)\right]} \nonumber \\[10pt]
&+ A_{P}(M'_{P})^{2}\cdot(T/2-t)^{2}\cdot\cosh{\left[M_{P}(T/2-t)\right]}~.
\end{align}
The first and second derivatives must be evaluated at $\Delta m=0$, where several simplifications occur:  the first derivative of the pion masses and amplitudes at $\Delta m=0$ vanishes
\begin{align}
M'_{\pi^{0}}\big|_{\Delta m=0} =  M'_{\pi^{+}}\big|_{\Delta m=0} =  A'_{\pi^{0}}\big|_{\Delta m=0} = A'_{\pi^{+}}\big|_{\Delta m=0} =0.
\end{align}
and one also has
\begin{align}
M_{\pi^{+}}\big|_{\Delta m=0} = M_{\pi^{0}}\big|_{\Delta m=0} \equiv M_{\pi},\quad A_{\pi^{+}}\big|_{\Delta m=0} = A_{\pi^{0}}\big|_{\Delta m=0} \equiv A_{\pi}~.
\end{align}
This implies that
\begin{align}
\label{M1_final}
\left[C''_{\pi^{+}\pi^{+}}(t) - C''_{\pi^{0}\pi^{0}}(t)\right]_{\Delta m=0} &= \left(A''_{\pi^{+}}-A''_{\pi^{0}}\right)\cosh{\left[M_{\pi}(T/2-t)\right]} \nonumber\\[10pt]
&+ \left( M''_{\pi^{+}}-M''_{\pi^{0}}\right)A_{\pi}\cdot(T/2-t)\cdot\sinh{\left[ M_{\pi}(T/2-t)\right]}~.
\end{align}
The isosymmetric pion mass, $M_{\pi}$, and amplitude, $A_{\pi}$, can be computed from the ground state of the isosymmetric pion correlator $C_{\pi\pi}^{\rm{isoQCD}}$, represented by the single connected diagram without any mass insertion, and computed with opposite values of the Wilson parameter $r$. We find it convenient to cancel the time dependence in the first term of Eq.~(\ref{M1_final}) by normalizing the expression to the charged pion correlator $C_{\pi\pi}^{\rm{isoQCD}}$, obtaining
\begin{align}
\frac{\left[C''_{\pi^{+}\pi^{+}}(t) - C''_{\pi^{0}\pi^{0}}(t)\right]_{\Delta m=0}}{C_{\pi\pi}^{\rm{isoQCD}}(t)} = \frac{A''_{\pi^{+}} - A''_{\pi^{0}}}{A_{\pi}} + (M''_{\pi^{+}}- M''_{\pi^{0}})\cdot(T/2-t)\cdot\tanh{\left[ M_{\pi}(T/2-t)\right]}~.
\end{align}
Given that at second order in $\Delta m$
\begin{align}
C_{\pi^{+}\pi^{+}}(t)-C_{\pi^{0}\pi^{0}}(t) &= \frac{1}{2}(\Delta m)^{2}\left[C''_{\pi^{+}\pi^{+}}(t) - C''_{\pi^{0}\pi^{0}}(t)\right]_{\Delta m=0} \\[10pt]
M_{\pi^{+}}-M_{\pi^{0}} &= \frac{1}{2}(\Delta m)^{2}\left[ M''_{\pi^{+}} - M''_{\pi^{0}}\right]~,
\end{align}
the charged/neutral pion mass difference at order $\mathcal{O}((\Delta m)^{2})$ can be extracted through a standard  ``effective slope''  analysis in time of the diagrams displayed in Eq.~(\ref{M1_expansion}), as it will be explained in the next section. \\

Let us now discuss the matrix element method and consider first the correlator $C_{P^{0}\pi^{0}}(t)$ in the physical basis. Inserting a complete set of eigenstates between the pion source and the isoscalar operator, one gets 
\begin{align}
C_{P^{0}\pi^{0}}(t) &= {\displaystyle \sum_{n}}\langle 0| P^{0}(t)|n\rangle\frac{1}{2E_{n}}\langle n| P^{\dag}_{\pi^{0}}(0)|0\rangle \nonumber \\[10pt]
&= \langle 0| P^{0}|\pi^{0}\rangle\frac{1}{M_{\pi^{0}}}\langle \pi^{0}| P^{\dag}_{\pi^{0}}|0\rangle e^{-M_{\pi^{0}}T/2}\cosh{\left[M_{\pi^{0}}(T/2-t)\right]} +\ldots 
\end{align}
where the dots correspond to terms that are exponentially suppressed w.r.t. the neutral pion contribution for large time separations $t\gg a, T-t \gg a$. Assuming the ground state dominance we have
\begin{align}
C_{P^{0}\pi^{0}}(t) &= \frac{Z_{P^{0}\pi^{0}}\cdot Z_{\pi^{0}\pi^{0}}}{M_{\pi^{0}}}e^{-M_{\pi^{0}}T/2}\cosh{\left[M_{\pi^{0}}(T/2-t)\right]}~,
\end{align}
where
\begin{align}
Z_{P^{0}\pi^{0}} = \langle 0| P^{0}|\pi^{0}\rangle.
\end{align}
As in the previous case, the matrix elements $Z_{P^{0}\pi^{0}}, Z_{\pi^{0}\pi^{0}}$ and the neutral pion mass $M_{\pi^{0}}$ are implicitly function of $\Delta m$. Therefore we can expand $C_{P^{0}\pi^{0}}$ at first order in $\Delta m$, in terms of the derivatives of $Z_{P^{0}\pi^{0}}, Z_{\pi^{0}\pi^{0}}$ and $M_{\pi^{0}}$. In this case, however, $Z_{P^{0}\pi^{0}}$ vanishes in the isosymmetric limit, because the neutral pion does not have an isoscalar component for $\Delta m=0$, and one gets
\begin{align}
C_{P^{0}\pi^{0}}(t) = \Delta m~ C'_{P^{0}\pi^{0}}(t)\big|_{\Delta m =0}~,    
\end{align}
where
\begin{align}
C'_{P^{0}\pi^{0}}(t)\big|_{\Delta m=0} = \frac{Z'_{P^{0}\pi^{0}}\cdot Z_{\pi\pi}}{M_{\pi}}e^{-M_{\pi}T/2}\cosh{\left[M_{\pi}(T/2-t)\right]}
\end{align}
and
\begin{align}
Z_{P^{0}\pi^{0}} = \Delta m\, Z'_{P^{0}\pi^{0}} + \mathcal{O}( (\Delta m)^{2}),\qquad Z_{\pi^{0}\pi^{0}}\big|_{\Delta m=0} \equiv Z_{\pi\pi}~.
\end{align}
In order to isolate the quantity we are interested in, namely $Z'_{P^{0}\pi^{0}}$, it is useful to normalize again over the isosymmetric charged pion correlator
\begin{align}
\label{eq:corrisoQCD}
C_{\pi\pi}^{\rm{isoQCD}}(t) = \frac{ |Z_{\pi\pi}|^{2}}{M_{\pi}}e^{-M_{\pi}T/2}\cosh{\left[M_{\pi}(T/2-t)\right]}~.
\end{align}
In this way we get
\begin{align}
\label{normalized_M2}
\frac{C'_{P^{0}\pi^{0}}(t)\big|_{\Delta m=0}}{C_{\pi\pi}^{\rm{isoQCD}}(t)} = \frac{Z'_{P^{0}\pi^{0}}}{Z_{\pi\pi}}~.
\end{align}
By performing a constant fit to the ratio of correlators in Eq.~(\ref{normalized_M2}), and by extracting  $Z_{\pi\pi}$ from $C_{\pi\pi}^{\rm{isoQCD}}(t)$, it is possible to determine $Z'_{P^{0}\pi^{0}}$, and thus $Z_{P^{0}\pi^{0}}$ at $\mathcal{O}(\Delta m)$.


\section{Numerical results}
\label{sec:Num_result}
We performed simulations on the cA211.30.32 ensemble generated by the ETMC~\cite{alexandrou2021ratio}, which has a spatial extent $L=32$ and aspect ratio $T/L = 2$. The number of gauge configurations that have been analyzed is  $N_{cfg} = 1232$. The ensemble corresponds to an higher-than-physical pion mass $M_{\pi}\simeq 260~{\rm MeV}$ with $M_{\pi}L \simeq 4.01$, and to a lattice spacing $a\simeq 0.095~{\rm fm} $. We made use of local sources to interpolate the pion fields,  and used one stochastic source per time in order to invert the Dirac operator. The computational cost of the simulation is of about $7\cdot 10^{4}~\rm{Core}\,\rm{Hours}$. \\

Relying on Eqs.~(\ref{l7_M1_definition}) and~(\ref{l7_M2_definition}), which are the basis for our mass and matrix element methods, we have built the following estimators to extract $\ell_{7}$ from the diagrams in Eqs.~(\ref{M1_expansion}) and~(\ref{M2_expansion}):
\begin{align}
\label{estimator_M1}
\bar{\ell}_{7}(t) &= \left(\frac{Z_{S}}{Z_{P}}\right)^{2}\cdot\frac{\hat{f}_{\pi}^{2}\hat{m}_{\ell}^{2}}{\hat{M}_{\pi}^{3}}\cdot \partial_{t}\left[\frac{ C_{MM}^{conn.}(t) - C_{MM}^{disc.}(t)}{C_{\pi\pi}^{\rm{isoQCD}}(t)}\right]\quad\,\,\,\,\,\,\,\,\,\,\, \left(\text{mass method}\right)~,\\[10pt]
\label{estimator_M2}
\bar{\ell}_{7}(t) &= -\left(\frac{Z_{S}}{Z_{P}}\right)\cdot\frac{ \hat{f}_{\pi}\hat{m}_{\ell}^{2}}{\hat{M}_{\pi}^{4}}\cdot \hat{Z}_{\pi\pi}\cdot\left[\frac{ C_{MEM}^{conn.}(t) - C_{MEM}^{disc.}(t)  }{ C_{\pi\pi}^{\rm{isoQCD}}(t)} \right]\quad\left(\text{matrix element method}\right)~.
\end{align}
In our twisted mass setup, the pion decay constant can be obtained from $\hat{Z}_{\pi\pi}$, which is the bare matrix element $\langle \pi^{0} | P_{\pi^{0}}^{\dag}|0\rangle$ extracted from $C_{\pi\pi}^{\rm{isoQCD}}$, using\footnote{Alternatively, the pion decay constant can be computed from the correlation function $C_{A^{0}\pi^{0}}(t)= \langle 0 | J_{5}^{0}(t) P^{\dag}_{\pi^{0}}(0)|0\rangle$, where $J_{5}^{0}$ is the zeroth component of the axial current. We checked that the two methods give similar values of $\hat{f}_{\pi}$, and that the resulting values of $\ell_{7}$ are the same within errors.}
\begin{align}
\hat{f}_{\pi} = 2\hat{m}_{\ell}\,\frac{\hat{Z}_{\pi\pi}}{\hat{M}_{\pi}\sinh{(\hat{M}_{\pi})}}~.    
\end{align}
The operator $-\partial_{t}$ in Eq.~(\ref{estimator_M1}), corresponds to the evaluation of the so-called effective slope $\delta m_{eff}(t)$ from the ratio of correlators $\delta C/C$,  which is defined through
\begin{align}
\label{eq:def_meff}
\delta m_{eff}(t) \equiv -\partial_{t} \frac{\delta C(t)}{C(t)} &= \frac{1}{F(T/2 -t ,M)}\left(\frac{\delta C(t)}{C(t)} -\frac{\delta C(t-1)}{C(t-1)}\right) ~ ,
\end{align}
where in our case
\begin{align}
\delta C(t) \equiv C_{MM}^{conn.}(t) - C_{MM}^{disc.}(t),\qquad C(t) \equiv C_{\pi\pi}^{isoQCD}(t)~.
\end{align}
In Eq.~(\ref{eq:def_meff}), $M$ is the ground state mass extracted from the correlator $C(t)$, $T$ is the temporal extent of the lattice and $F(x,M)$ is an analytical factor given by
\begin{align}
F(x,M) &= x\tanh{(Mx)} - (x+1)\tanh{(M(x+1))} ~ .
\end{align}
In the large time limit $t \gg a, (T-t) \gg a$, both estimators in Eqs.~(\ref{estimator_M1}) and~(\ref{estimator_M2}) tend to $\ell_{7}$.\\
\begin{table}
    \centering
    \begin{tabular}{c| c c c c c}
    \hline
         & $\hat{m}_{\ell}$ & $\hat{f}_{\pi}$ & $\hat{M}_{\pi}$ &  $Z_{P}/Z_{S}$   \\
    cA211.30.32     &  0.0030 & 0.06674 (15)  & 0.12530 (16) & 0.726 (3) \\
    \hline
    \end{tabular}
    \caption{List of the input parameters entering the determination of $\ell_{7}$ as obtained on the cA211.30.32 ensemble. $\hat{m}_{\ell}$ is the bare light lattice quark mass.}
    \label{tab1}
\end{table}
\begin{figure}
\centering
    \includegraphics[scale=0.58]{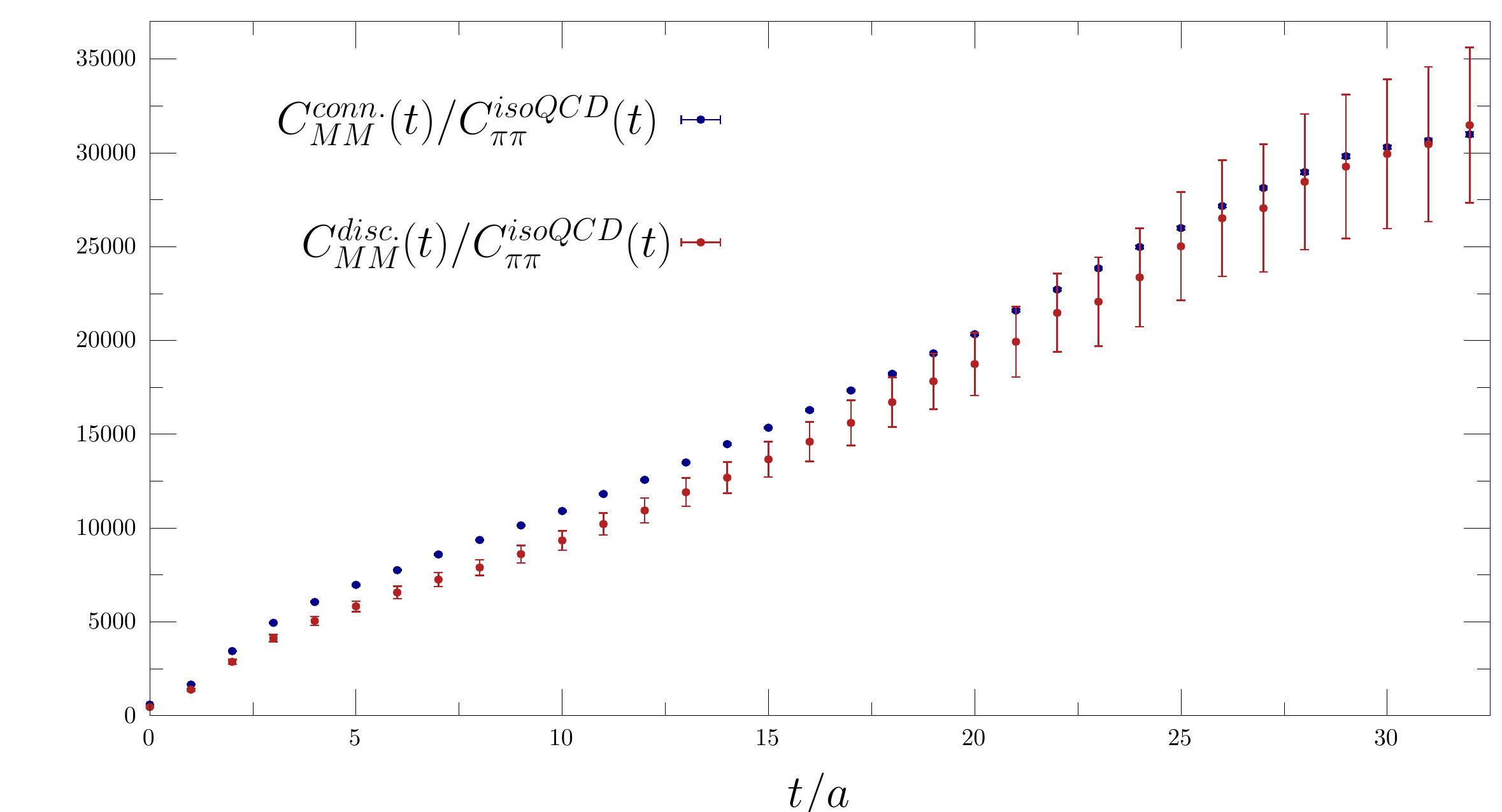}
    \caption{\label{fig1}Comparison between the connected diagram $C_{MM}^{conn.}(t)$ and the disconnected diagram $C_{MM}^{disc.}(t)$ contributing to $C_{\pi^{+}\pi^{+}}(t)-C_{\pi^{0}\pi^{0}}(t)$ and normalized over the isosymmetric charged pion correlator $C_{\pi\pi}^{\rm{isoQCD}}(t)$.}
    
    \vspace{1cm}
    \includegraphics[scale=0.58]{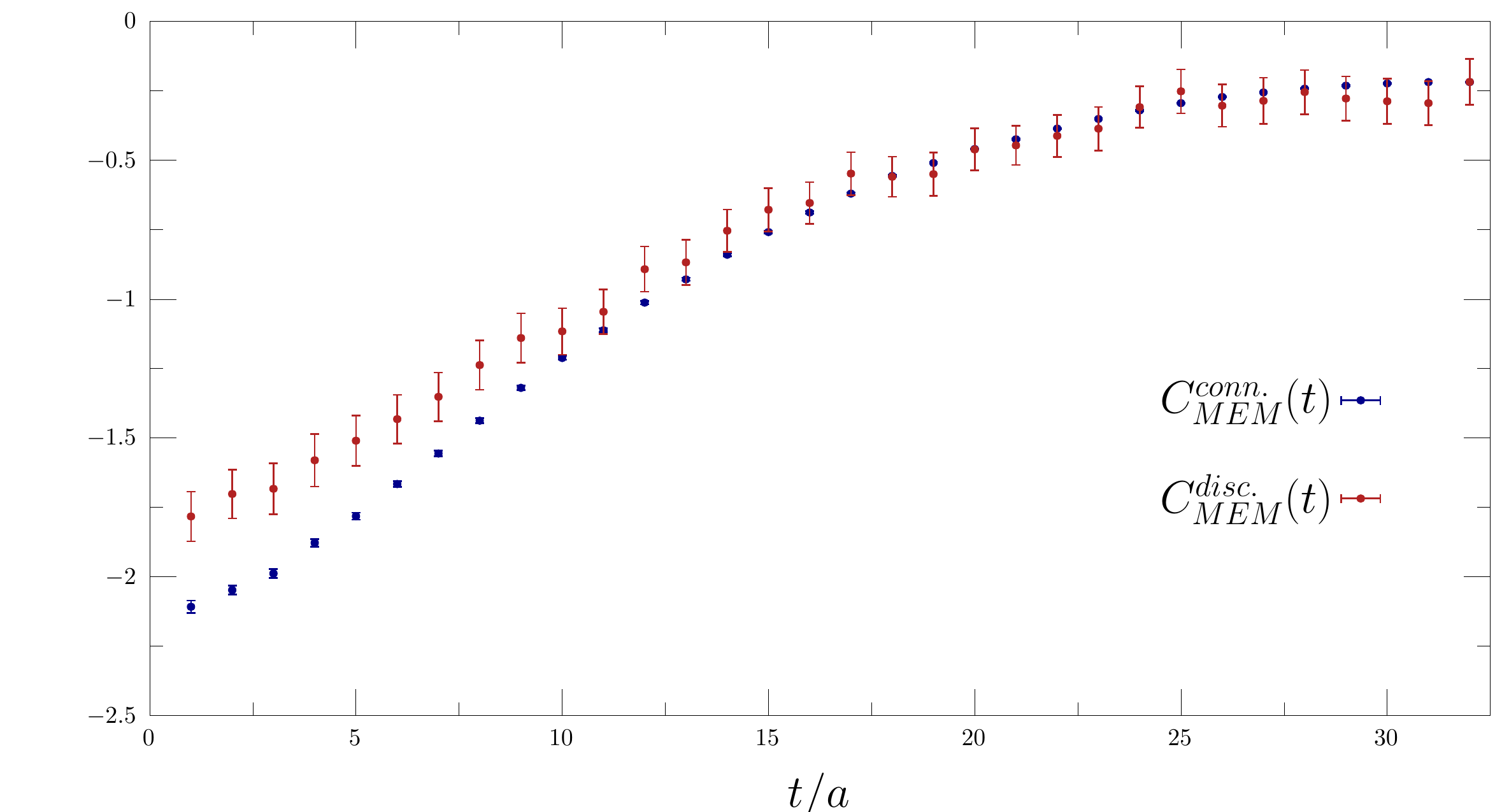}
       \caption{\label{fig2} Comparison between the connected diagram $C_{MEM}^{conn.}(t)$ and the disconnected diagram $C_{MEM}^{disc.}(t)$ contributing to $C_{P^{0}\pi^{0}}(t)$.}
  
\end{figure}

\begin{figure}
    \centering
    \includegraphics[scale=0.7]{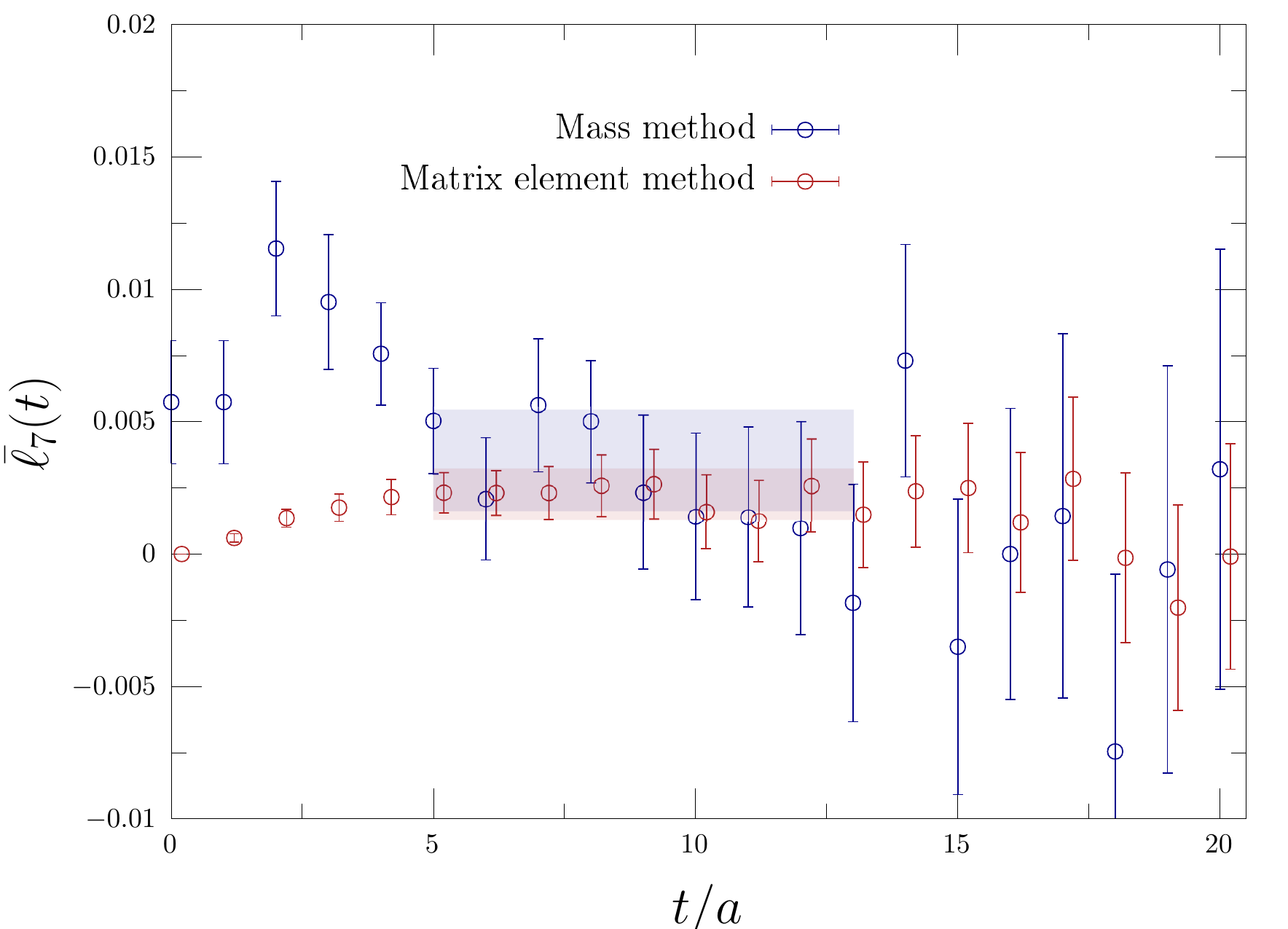}
    \caption{Determination of $\ell_{7}$ on the cA211.30.32 ensemble using both the mass method and the matrix element method. The semi-transparent bands correspond to the result of a constant fit in the time interval $[5,13]$.}
    \label{fig3}
\end{figure}

In Tab.~\ref{tab1} we collected the values of the input parameters that have been used for the determination of $\ell_{7}$ on the cA211.30.32 ensemble. The ratio between the RCs $Z_{S}$ and $Z_{P}$ has been computed using
\begin{align}
\frac{Z_{S}}{Z_{P}} = \frac{\hat{Z}_{\pi\pi}}{\hat{Z}_{\pi\pi}^{OS}}~,
\end{align}
where $\hat{Z}_{\pi\pi}^{OS}$ is the bare matrix element $\langle \pi^{0} | P_{\pi^{0}}^{\dag}|0\rangle$ extracted from the Osterwalder-Seiler pion correlator, i.e. from the single connected diagram in which the quark and antiquark propagators are computed with the same value of the Wilson parameter $r$. In Fig.~\ref{fig1} we show our determination of the diagrams $C_{MM}^{conn.}(t)$ and $C_{MM}^{disc.}(t)$ normalized over the isosymmetric charged pion correlator $C_{\pi\pi}^{\rm{isoQCD}}(t)$, while  in Fig.~\ref{fig2} we show our determination of $C_{MEM}^{conn.}(t)$ and $C_{MEM}^{disc.}(t)$. As the figures show, for both mass and matrix element methods, the signal of $\ell_{7}$ comes from a  large cancellation between the connected and the disconnected contributions. This makes the evaluation of $\ell_{7}$ a non-trivial task, since a very good precision on both diagrams is needed to ensure that the difference is not dominated by the statistical noise. In this respect, the use of the RTM scheme turns out to be crucial (see Ref.~\cite{Fr:2021} for more details on this improvement).
In Fig.~\ref{fig3}, we show our determination of $\bar{\ell}_{7}(t)$ as obtained from both mass and matrix element methods. As can be seen, the matrix element method seems to perform better, giving a more precise result w.r.t. the mass method. In both cases, the signal disappears after a time separation $t/a \sim 15$, and $\ell_{7}$ can be extracted only at smaller times. We decided to fit both estimators in the time interval $[5,13]$. We obtain in this way
\begin{align}
\label{eq:l7_final_M1}
\ell_{7} &= 3.5(2.0)\times 10^{-3},\qquad \left(\text{mass method}\right)~, \\[10pt]
\label{eq:l7_final_M2}
\ell_{7} &= 2.3(1.0)\times 10^{-3},\qquad \left(\text{matrix element method}\right)~.
\end{align}

Even if our analysis is limited to a single value of the lattice spacing and to a single pion mass $M_{\pi}\simeq 260~{\rm MeV}$, we can give a rough estimate of the systematic error due to the missing chiral and continuum extrapolations. The two results of Eqs.~(\ref{eq:l7_final_M1}) and~(\ref{eq:l7_final_M2}), obtained from the mass and the matrix element methods, are affected in principle by different lattice artifacts, and the deviation among their central values can be taken as a first (likely conservative) estimate of the $\mathcal{O}(a^2)$ effects. Instead, for the light quark mass dependence, it is reasonable to assume that the impact of the extrapolation $m_{\ell}\to 0$ is negligible as compared to the statistical uncertainty affecting our determination, given that in ChPT the presence of a non-zero light quark mass $m_{\ell}$ corresponds to a NNLO correction to the formulae in Eqs.~(\ref{eq:mpi_l7}) and~(\ref{eq:Zp0pi_l7}). Our (conservative) estimate of the value of $\ell_{7}$ is
\begin{align}
\label{eq:final_l7}
\ell_{7} = 2.5(1.3)_{stat.}(0.5)_{syst.}\times 10^{-3} = 2.5(1.4)\times 10^{-3}~,
\end{align}
where the central value and the error estimate have been obtained from the two determinations of Eqs.~(\ref{eq:l7_final_M1}) and~(\ref{eq:l7_final_M2}) making use of Eqs.~(38)-(43) of Ref.~\cite{alexandrou2021quark}. \\

Our determination can be compared with the phenomenological estimate given in Ref.~\cite{di_Cortona_2016}
\begin{align}
\ell_{7}^{ph.} = 7(4) \times 10^{-3} ~,
\end{align}
and with the global-fit based result obtained by the RBC-UKQCD Collaboration $\ell_{7}= 6.5(3.8)\times 10^{-3}$~\cite{Boyle_2016}. Our result is in agreement but significantly improves both estimates, and shows the effectiveness of the RM123 approach to determine $\ell_{7}$.


\section{Conclusions}
\label{sec:Conclusions}
In this paper we showed that it is possible to determine directly from lattice QCD calculations the $\rm{SU}(2)$ ChPT low-energy costant $\ell_{7}$ which parametrizes QCD isospin-breaking effects in the chiral Lagrangian at NLO and that is crucial for several phenomenological analyses. Our strategy is based on the the RM123 approach~\cite{de_Divitiis_2012, Giusti:2017dmp}, which allows to evaluate isospin-breaking effects perturbatively in the up-down quark mass difference $m_{u}-m_{d}$. In addition, in order to increase the precision of the lattice correlators involved in the calculation we made use of the recently proposed \textit{rotated twisted-mass} (RTM) scheme~\cite{Fr:2021}. \\

To determine $\ell_{7}$ we explored two strategies. The first one, the mass method, is based on the computation of the charged/neutral pion mass difference $M_{\pi^{+}}- M_{\pi^{0}}$, whose second derivative in $m_{u}-m_{d}$, evaluated at the isosymmetric point $m_{u}=m_{d}$, is directly proportional to $\ell_{7}$. The second strategy, the matrix element method, allows to compute $\ell_{7}$ from the slope in $m_{u}-m_{d}$ of the coupling $Z_{P^{0}\pi^{0}}$ between the neutral pion and the isoscalar operator $P^{0}$, which is a pure isospin breaking effect. The two methods give rise to consistent results, but we find that the matrix element method  displays an higher statistical accuracy w.r.t. the mass method. The comparison of our determination with existing results reveals substantial agreement, although the difference is slightly larger than one standard deviation. It is however important to remind that our results are obtained at a single value of the lattice spacing and with a pion mass $M_{\pi} \simeq 260~{\rm MeV}$. Therefore, our result should be understood as a proof-of-principle calculation, showing the feasibility of a direct determination of $\ell_{7}$ from lattice QCD calculations. In the future we plan to perform simulations at different values of the lattice spacing and explore different light quark masses in order to perform a reliable extrapolation towards the continuum and chiral limit. A possibility would be to extend the calculation to the other ensembles produced by the ETM Collaboration with Wilson-clover TM fermions, or alternatively, one could consider other lattice discretizations.


\section{Acknowledgement}
We thank L. Di Luzio, G. Piazza and C. Tarantino for useful discussions, and all members of ETMC for the most enjoyable collaboration. We acknowledge CINECA for the provision
of CPU time under the specific initiative INFN-LQCD123 and IscrB\_S-EPIC. F.S. G.G and S.S. are supported by
the Italian Ministry of University and Research (MIUR) under grant PRIN20172LNEEZ. F.S. and G.G are supported by INFN under GRANT73/CALAT.

\bibliographystyle{ieeetr}
\bibliography{references}

\begin{thebibliography}{10}

\bibitem{Gasser:1983yg}
J.~Gasser and H.~Leutwyler, ``{Chiral Perturbation Theory to One Loop},'' {\em
  Annals Phys.}, vol.~158, p.~142, 1984.

\bibitem{GASSER1985465}
J.~Gasser and H.~Leutwyler, ``Chiral perturbation theory: Expansions in the
  mass of the strange quark,'' {\em Nuclear Physics B}, vol.~250, no.~1,
  pp.~465--516, 1985.

\bibitem{di_Cortona_2016}
G.~G. di~Cortona, E.~Hardy, J.~P. Vega, and G.~Villadoro, ``The {QCD} axion,
  precisely,'' {\em Journal of High Energy Physics}, vol.~2016, Jan 2016.

\bibitem{diluzio2021axion}
L.~D. Luzio, G.~Martinelli, and G.~Piazza, ``Axion hot dark matter bound,
  reliably,'' 2021.
\newblock e-Print: 2101.10330 [hep-ph].

\bibitem{Boyle_2016}
P.~A. Boyle, N.~H. Christ, N.~Garron, C.~Jung, A.~Jüttner, C.~Kelly, R.~D.
  Mawhinney, G.~McGlynn, D.~J. Murphy, S.~Ohta, and et~al., ``Low energy
  constants of {$\rm{SU}(2)$} partially quenched chiral perturbation theory
  from {$N_{f} =2+1$} domain wall {QCD},'' {\em Physical Review D}, vol.~93,
  no.~5.

\bibitem{de_Divitiis_2012}
G.~M. de~Divitiis, P.~Dimopoulos, R.~Frezzotti, V.~Lubicz, G.~Martinelli,
  R.~Petronzio, G.~C. Rossi, F.~Sanfilippo, S.~Simula, and et~al., ``Isospin
  breaking effects due to the up-down mass difference in lattice {QCD},'' {\em
  Journal of High Energy Physics}, no.~4, 2012.

\bibitem{Giusti:2017dmp}
D.~Giusti, V.~Lubicz, C.~Tarantino, G.~Martinelli, F.~Sanfilippo, S.~Simula,
  and N.~Tantalo, ``{Leading isospin-breaking corrections to pion, kaon and
  charmed-meson masses with Twisted-Mass fermions},'' {\em Phys. Rev. D},
  vol.~95, no.~11, p.~114504, 2017.

\bibitem{Fr:2021}
R.~Frezzotti, G.~Gagliardi, V.~Lubicz, F.~Sanfilippo, and S.~Simula, ``Rotated
  twisted-mass: a convenient regularization scheme for isospin breaking {QCD}
  and {QED} lattice calculations,'' 2021.
\newblock e-Print: 2106.07107 [hep-lat].

\bibitem{alexandrou2021ratio}
C.~Alexandrou, S.~Bacchio, G.~Bergner, P.~Dimopoulos, J.~Finkenrath,
  R.~Frezzotti, M.~Garofalo, B.~Kostrzewa, G.~Koutsou, P.~Labus, F.~Sanfilippo,
  S.~Simula, M.~Ueding, C.~Urbach, and U.~Wenger, ``Ratio of kaon and pion
  leptonic decay constants with {$N_f = 2 + 1 + 1$} {Wilson-clover}
  twisted-mass fermions,'' 2021.
\newblock e-Print: 2104.06747 [hep-lat].

\bibitem{alexandrou2021quark}
C.~Alexandrou, S.~Bacchio, G.~Bergner, M.~Constantinou, M.~D. Carlo,
  P.~Dimopoulos, J.~Finkenrath, E.~Fiorenza, R.~Frezzotti, M.~Garofalo,
  K.~Hadjiyiannakou, B.~Kostrzewa, G.~Koutsou, K.~Jansen, V.~Lubicz,
  M.~Mangin-Brinet, F.~Manigrasso, G.~Martinelli, E.~Papadiofantous,
  F.~Pittler, G.~C. Rossi, F.~Sanfilippo, S.~Simula, C.~Tarantino, A.~Todaro,
  C.~Urbach, and U.~Wenger, ``Quark masses using twisted mass fermion gauge
  ensembles,'' 2021.
\newblock e-Print: 2104.13408 [hep-lat].

\bibitem{Frezzotti_2004}
R.~Frezzotti and G.~Rossi, ``Twisted-mass lattice {QCD} with mass
  non-degenerate quarks,'' {\em Nuclear Physics B - Proceedings Supplements},
  vol.~128, p.~193–202, Feb 2004.

\end{thebibliography}

\end{document}